\documentclass[a4]{aa}
\usepackage[dvips]{graphics}

\begin{document}

\title{The Calar Alto lunar occultation program:
update and new results\thanks{
Based on observations collected at TIRGO (Gornergrat, Switzerland), and at Calar
Alto (Spain). TIRGO is operated by CNR--CAISMI Arcetri, Italy. Calar Alto is
operated by the German--Spanish Astronomical Center. 
Table~5 is only available
 in electronic form
at the CDS via anonymous ftp to cdsarc.u-strasbg.fr (130.79.128.5)
or via http://cdsweb.u-strasbg.fr/cgi-bin/qcat?J/A+A/
} 
}

\author{A. Richichi\inst{1}
\and O. Fors\inst{2}\fnmsep\inst{3}
\and M. Merino\inst{2}
\and X. Otazu\inst{4}
\and J. N\'u\~nez\inst{2}\fnmsep\inst{3}
\and A. Prades\inst{5}
\and U. Thiele\inst{6}
\and D. P\'erez-Ram\'\i rez\inst{7}
\and F.J. Montojo\inst{8}
}

\institute{
European Southern Observatory,
Karl-Schwarzschild-Str. 2, D-85748 Garching bei M\"unchen, Germany
\and
Departament d'Astronomia i Meteorologia, Universitat de  Barcelona,
Mart\'{\i} i Franqu\'es 1, D-08028 Barcelona, Spain
\and
Observatori Fabra, Cam\'{\i} de l'Observatori s/n, D-08035 Barcelona, Spain
\and
Computer Vision Center, UAB, Bellaterra, Spain
\and
Escola Universitaria Polit\`ecnica de Barcelona,
Universitat Polit\`ecnica de Catalunya, Barcelona, Spain
\and
Calar Alto Observatory, Almer\'{\i}a, Spain
\and
Universidad de Ja\'en, Dpto. de F\'\i sica, Campus Las 
Lagunillas s/n, E-23071, Ja\'en, Spain
\and
Real Instituto y Observatorio de la Armada, San Fernando, Spain
}
\offprints{A. Richichi, \email{arichich@eso.org}}
\date{Received / Accepted }
\titlerunning{}

\abstract{
We present an update of the lunar occultation program which
is routinely carried out in the near-IR at the 
Calar Alto Observatory. A total of 350 events were
recorded since our last report (Fors et al.~\cite{fors04}).
In the course of eight runs we have observed, among others,
late-type giants, T-Tauri stars, and infrared sources.
Noteworthy
was a passage of the Moon close to the galactic center, which
produced a large number of events during just a few hours in July 2004.
Results include the determinations of the angular diameter
of \object{RZ Ari}, and the projected separations and
brightness ratios for
one triple and 13 binary stars,
almost all of which representing first time detections.
Projected separations range from $0\farcs09$ to $0\farcs007$.
We provide a quantitative 
analysis of the performance achieved in our observations in terms
of angular resolution and sensitivity,
which reach about $0\farcs003$ and $K \approx8.5$\,mag, 
respectively.
We 
also present a statistical discussion of our sample, and
in particular of the frequency of detection of binaries among
field stars.
}
\maketitle

\section{Introduction}
Among the various methods to achieve high-angular resolution beyond the
limits set by atmospheric turbulence and by the telescope diffraction,
lunar occultations (LO) stand out for their relative simplicity.
LO have been used intensively for investigations of stellar sources
for a few decades, and have led to systematic determinations of angular
diameters and binary stars. A recent census of LO results
was provided by Richichi et al.~\cite{CHARM2}.  
In recent years attention and
resources are shifting to methods such as long-baseline interferometry (LBI),
which offer almost complete freedom of choice in the targets and
times of observation, although with much more demanding technical
requirements. However, in our opinion LO can still offer a significant
scientific contribution with relatively small effort, and attempts
to continue routine LO observations should be encouraged.
One important advancement in this area has been 
the recent availability of all-sky near-infrared surveys, such as 
2MASS (Cutri et al.~\cite{cutri03}) and DENIS (Paturel et al.~\cite{paturel03}). Such catalogues have increased by
almost an order of magnitude the
number of predictions for events observable with 1\,m-class telescopes
and above.

In this paper, we provide an overview of the data we have accumulated
since our last publication on the subject (Fors et al.~{\cite{fors04},
F04 hereafter),
adding 350 LO events. We present new approaches in data
analysis, suited to the automated reduction of large volumes
of LO, and we provide details on 11 new binaries and one triple star, 
2 known binaries,
and one angular diameter determinations.
We also discuss the statistics of our sample, with some 
considerations on the detection of binaries in random observations
of field stars.

\section{Observations, data handling and data reduction}\label{data}
Most of the observations reported here
were carried out with the 1.5\,m telescope of the Observatorio 
Astron\'omico Nacional in Calar Alto (Spain). On two occasions, we used 
the 3.5\,m and
2.2\,m telescopes  
of the Centro Astron\'omico Hispano-Alem\'an, 
located at the same site.
Among the present results
we have included also two earlier observations
obtained at the TIRGO 1.5\,m telescope, for which a detailed re-analysis
has shown a positive binary detection. The above telescopes, the
associated instrumentation  and the filter bandpasses used for
LO work have been described elsewhere 
(Richichi et al.~\cite{richichi96}, F04).
Concerning the K-filter of the MAGIC cameras used at 1.5\,m and
2.2\,m telescopes of Calar Alto, we have determined an improved,
accurate transmission curve at operating temperature, and applied
it in all relevant cases. For this purpose we have used the same MAGIC
camera in the laboratory, taking exposures with and without
the filter, using the same resin-replica grism used for
astronomical observations at liquid nitrogen temperature.
As a light source we employed
a source without significant emission in the J band, thus
avoiding contamination of short-wavelength light from
a different order.
For the observation of the very bright star \object{RZ Ari} we
employed a narrow band filter, with $\lambda_{\rm 0}=2.26\mu$m and
$\Delta \lambda=0.06\mu$m.

Observations were carried out during eleven observing runs
at the Calar Alto Observatory over a period of two years,
as detailed in
Table~\ref{table_runs}.
On average, each run consisted of a few nights allocated
in periods of crescent Moon close to full phase,
in order to maximize the number of occultations of
field stars and observe disappearances rather than
reappearances. On two occasions (runs G and H), very
short runs were allocated to follow up  passages
of the Moon close to the galactic center and over
the Taurus star-forming regions, respectively.
Note that three runs were completely devoid of results
due to weather. The two earlier results from TIRGO are
collectively grouped as run A, although their respective observations
were collected at different dates in 2001.
Details of the 350 recorded events and 
the characteristics of the corresponding objects can be found
in Table~5, available only on-line.
Here we show an excerpt in Table~\ref{table_list}, which
contains the details of the sources explicitly mentioned
either for a positive result or for other comments.
The format of this
table is similar to the one used in F04.
In column (3), the codes T, CB and CC are for the TIRGO telescope
equipped with a fast InSb photometer, 
and for the Calar Alto 1.5,m and 2.2\,m telescopes equipped
with MAGIC cameras, respectively.

\begin{table}
\caption[]{Observing runs}
\label{table_runs}
\begin{tabular}{cccrr}
\hline 
\hline 
\multicolumn{1}{c}{Run}&
\multicolumn{1}{c}{Telescope}&
\multicolumn{1}{c}{Dates}&
\multicolumn{1}{c}{Nights}&
\multicolumn{1}{c}{\# LO}\\
\hline 
   A & TIRGO 1.5m & Oct, Nov 01 &          2 &          2 \\
   B &    CA 1.5m &     Feb 03 &  	5 &	     0 \\
   C &    CA 1.5m &     Nov 03 &  	5 &	     9 \\
   D &    CA 1.5m &     Dec 03 &  	5 &	     0 \\
   E &    CA 1.5m &     Feb 04 &  	6 &	    29 \\
   F &    CA 1.5m &     Mar 04 &  	7 &	     3 \\
   G &    CA 2.2m &     Jul 04 &        0.5 &	    54 \\
   H &    CA 3.5m &     Oct 04 &  	1 &	     0 \\
   I &    CA 1.5m &     Nov 04 &  	6 &	    45 \\
   J &    CA 1.5m &     Dec 04 &  	5 &	     7 \\
   K &    CA 1.5m &     Jan 05 &  	5 &	   105 \\
   L &    CA 1.5m &     Feb 05 &  	5 &	    96 \\
\cline{3-5}
     &            &      Total &       53.5 & 350 \\
\hline                                                                      
\hline                                                                      
\end{tabular}                                                               
\\                                                                          
\end{table}

\begin{table*}
\caption{List of selected occultation events and of the circumstances of their observation\label{table_list}}
\begin{tabular}{lccrrrrllc}
\hline 
\hline 
\multicolumn{1}{c}{(1)}&
\multicolumn{1}{c}{(2)}&
\multicolumn{1}{c}{(3)}&
\multicolumn{1}{c}{(4)}&
\multicolumn{1}{c}{(5)}&
\multicolumn{1}{c}{(6)}&
\multicolumn{1}{c}{(7)}&
\multicolumn{1}{c}{(8)}&
\multicolumn{1}{c}{(9)}&
\multicolumn{1}{c}{(10)}\\
\multicolumn{1}{c}{Source}&
\multicolumn{1}{c}{Date}&
\multicolumn{1}{c}{Tel.+}&
\multicolumn{1}{c}{D}&
\multicolumn{1}{c}{$\Delta $t }&
\multicolumn{1}{c}{$\tau $ }&
\multicolumn{1}{c}{V}&
\multicolumn{1}{c}{K}&
\multicolumn{1}{c}{Sp.}&
\multicolumn{1}{c}{Dist.}\\
&
\multicolumn{1}{c}{UT}&
\multicolumn{1}{c}{detector}&
\multicolumn{1}{c}{$\arcsec $}&
\multicolumn{1}{c}{ms}&
\multicolumn{1}{c}{ms}&
\multicolumn{1}{c}{mag}&
\multicolumn{1}{c}{mag}&
\multicolumn{1}{c}{}&
 pc
\\
\hline 
\object{SAO 164567} & 25-10-01 & T & 21 & 3.0 & 3.4 & 7.4 & 3.2 & K5III & 278 \\
\object{SAO 110325} & 28-11-01 & T & 21 & 2.0 & 2.4 & 6.4 & 4.1 & K0 & 147 \\
\object{SAO 80310} & 03-03-04 & CB & 7 & 8.5 & 3.0 & 6.9 & 5.6 & F8 & 35 \\
\object{SAO 80764} & 01-04-04 & CB & 7 & 8.4 & 3.0 & 7.8 & 4.0 & K2 & 1429 \\
\object{SAO 185661} & 28-07-04 & CC & 5 & 8.4 & 3.0 & 9.9 & 5.9 & K5 &  \\
\object{IRC -30319} & 28-07-04 & CC & 5 & 8.4 & 3.0 & 8.8 & 1.8 & K2 &  \\
\object{17454891-2809333} & 28-07-04 & CC & 5 & 8.3 & 3.0 &  & 6.1 &  &  \\
\object{SAO 164601} & 18-11-04 & CB & 7 & 8.6 & 3.0 & 6.2 & 5.7 & A0m... & 110\\
\object{SAO 165154} & 19-11-04 & CB & 7 & 8.4 & 3.0 & 9.0 & 6.2 & K1III &  \\
\object{SAO 109617} & 22-11-04 & CB & 7 & 8.4 & 3.0 & 8.2 & 5.5 & K2 & 21 \\
\object{SAO 110089} & 23-11-04 & CB & 7 & 8.4 & 3.0 & 8.5 & 6.7 & K0 & 47 \\
\object{SAO 92659} & 23-11-04 & CB & 7 & 8.5 & 3.0 & 5.9 & 5.1 & F2Vw & 43 \\
\object{RZ Ari} & 18-01-05 & CB & 7 & 8.4 & 3.0 & 5.8 & -0.9 & M6III & 124 \\
\object{SAO 76214} & 19-01-05 & CB & 7 & 8.5 & 3.0 & 8.2 & 5.4 & K0 &  \\
\object{LH 98-106} & 19-01-05 & CB & 7 & 8.5 & 3.0 & 7.3 & 6.0 & F5 & 37 \\
\object{DL Tau} & 20-01-05 & CB & 7 & 8.4 & 3.0 & 13.6 & 8.0 & GV:e... &  \\
\object{GN Tau} & 20-01-05 & CB & 7 & 8.5 & 3.0 & 15.1 & 8.1 & M2.5 &  \\
\object{Elias 3-18} & 20-01-05 & CB & 7 & 8.5 & 3.0 &  &  & B5 &  \\
\object{ITG 31} & 20-01-05 & CB & 7 & 8.5 & 3.0 & 9.1 & 5.2 & K0 & 565 \\
\object{LkHA 332} & 21-01-05 & CB & 7 & 8.4 & 3.0 & 14.7 & 7.9 & K5 &  \\
\object{IRAS 04395+2521} & 21-01-05 & CB & 7 & 8.5 & 3.0 &  & 5.5 &  &  \\
\object{04440885+2540333} & 21-01-05 & CB & 7 & 8.6 & 3.0 & & 6.9 &  &  \\
\object{05415664+2707323} & 22-01-05 & CB & 7 & 8.5 & 3.0 & & & &  \\
\object{SAO 78540} & 23-01-05 & CB & 7 & 8.6 & 3.0 & 6.9 & 5.3 & G0 & 36 \\
\object{HD 283610} & 16-02-05 & CB & 7 & 8.5 & 3.0 & 9.6 & 5.4 & K5III &  \\
\object{04264187+2500314} & 17-02-05 & CB & 7 & 8.4 & 3.0 &  & 6.7 &  &  \\
\object{SAO 77000} & 17-02-05 & CB & 7 & 8.4 & 3.0 & 9.1 & 5.4 & G5 & 244 \\
\hline                                                                      
\hline 
\end{tabular}                                                               
\\                                                                          
\end{table*}

The availability of near-IR all-sky surveys has represented
a major step forward in the possibilities of LO investigations,
especially at medium and large-sized telescopes. Roughly speaking,
a 1.5\,m telescope equipped with an InSb fast photometer can
record LO events with millisecond time sampling and signal-to-noise
ratio (SNR) above unity for sources having magnitudes K$\la$7
(Richichi et al. \cite{richichi96}). An instrument based on the fast readout
of a subwindow of an array detector can add more than one
magnitude in sensitivity (F04). Moving to larger
telescopes brings gain in sensitivities which are essentially,
for moderate lunar phases and faint sources, proportional to
the area of the telescope (Richichi \cite{richichi96}). While these guidelines
are obviously strongly dependent on a number of instantaneous
parameters, they show that LO observations can be adequately
recorded on sources with magnitudes as faint as K$\approx$10.
Until recently, no comprehensive coverage of the near-IR sky
was available. The Two Micron Sky Survey (TMSS, 
Neugebauer \& Leighton~\cite{neugebauer69})
was incomplete in declination and only extended to K$\la$3. In the
past, LO 
predictions were compiled by the present authors using a variety
of other catalogues. Even a very rich run would consist of
about 10-20 sources per night at most.
We have now implemented the 2MASS survey (Cutri et al.~\cite{cutri03}) in our
predictions, and the number of events observable has jumped
up by very large factors. A typical night would offer in
excess of 100 sources close to maximum lunar phase. Even more
dramatic are the improved conditions for special events. For example,
on the occasions of passages of the Moon in crowded regions
near the Galactic Center 
(see Sect.\ref{other_interest}), 
thousand of events would be easily accessible
to a medium-sized telescope over few hours.
Table~\ref{table_gc} lists some statistics for such
events in the near future.
In this regime, the number of events effectively
observable will depend on the overheads of telescope pointing,
instrument operation and data storing.

\begin{table}
\caption[]{Statistics of three passages of the Moon
in the Galactic Center (GC) region in 2006, as
predicted for the Calar Alto Observatory
to the limit $K\leq8.5$ mag.}
\label{table_gc}
\begin{tabular}{lrrcc}
\hline
\hline
Date & Start & End & Minimum & Number of\\
  &  (UT)    &   (UT) & GC approach  & LO\\
\hline
11~June & 21:01 & 02:46 & $1\fdg95$ & 2899 \\
9~July & 20:31 & 01:48 & $4\fdg74$ & 1586 \\
5~August & 19:03 & 23:29 & $0\fdg88$ & 2315 \\
\hline
\hline
\end{tabular}
\end{table}

Of course, the  increase in the number of observed events is
not reflected linearly in the number of results, such as the
positive detection of field binaries, mainly because the majority
of the events will be faint and offer limited dynamic range. We will
return to this point in Sect.~\ref{performance}. The increase in the sheer
number of observed events, on the other hand, implies a significant
load of data inspection and analysis, particularly since data
from IR arrays are substantially more demanding than those
from photometers. 
This has prompted us to handle the bulk of raw
data by means of automated processing. A new reduction pipeline was designed and
implemented for the automated generation of preliminar lightcurve fits, which
are then improved interactively. In particular, we concentrated our
effort in two different areas. On one hand, we performed a
comparative study of different algorithms of light curve
extraction, such as aperture
photometry, Gaussian profile fit, object detection based on segmentation
analysis and subtraction of fixed number of faintest to brightest pixels. 
The latter was found to offer the best performance over the range of
SNR present in our data sets, using 30 and 15 pixels for the
extraction of the background and
star signals, respectively. 
On the other hand, a new algorithm was developed to estimate
automatically the lightcurves parameters (occultation time, stellar
and background intensity). A particular wavelet transform of the lightcurve was
chosen for this purpose, as it was capable of isolating the desired frequency
signature while preserving the temporal information. The
algorithm showed great robustness even in worst SNR conditions. This work
will be described in detail in a separate paper.

The data were analyzed by means of various methods,
as already described in Richichi et al.
(\cite{bina6}, \cite{diam6}) and references therein. The main
engine for data analysis is based on
a model-dependent least squares method (LSM).
Free parameters include the stellar intensity, the rate of the event, the
intensity of the background and its time drift.  For single stars
another parameter is 
the angular diameter, and additionally for binary
stars the  angular diameter of the companion, the projected separation and the
brightness ratio are included.
Spurious frequencies due to
pick-up of mains power and other effects may be present occasionally, and
can be digitally filtered. 
Relatively slow, random fluctuations of the background (due to thin
cirrus and lunar halo) and of the stellar intensity (due to image motion and
scintillation), can be fitted and accounted for 
by means of Legendre polynomials as described
in the above mentioned papers.
Another approach is to 
use a model-independent method (CAL, Richichi \cite{richichi89}), 
which is
particularly suited for the detection or confirmation 
of companions at very small separations. This method is also 
of great advantage in cases when the source may not be a
simple circular disk, or in the presence
of extended circumstellar emission.

\section{Results}\label{results}

The stars for which a positive result could be obtained are listed in
Table~\ref{table_results}, using the same format already used 
in F04.
In summary, the columns list the absolute value of the fitted linear rate of
the event V, its deviation from the predicted rate V$_{\rm{t}}$,
the local lunar limb slope $\psi$, the position and
contact angles, the signal--to--noise ratio (SNR). For binary detections, the
projected separation and the brightness ratio are given, while for 
\object{RZ Ari}
the angular diameter $\phi_{\rm UD}$ is reported, under the assumption of
a uniform stellar disc. 
All angular quantities are computed from the fitted
rate of the event. Only in 
\object{2MASS 04264187 +2500314} we were not
able to reliably fit a rate, due to the low SNR.
For this source, the 
predicted values
are listed in parentheses.
Note that in the tables of this paper the 2MASS prefix is 
omitted.

\begin{table*}
\caption{Summary of results\label{table_results}}
\begin{tabular}{lcrrrrrrrr}
\hline 
\hline 
\multicolumn{1}{c}{(1)}&
\multicolumn{1}{c}{(2)}&
\multicolumn{1}{c}{(3)}&
\multicolumn{1}{c}{(4)}&
\multicolumn{1}{c}{(5)}&
\multicolumn{1}{c}{(6)}&
\multicolumn{1}{c}{(7)}&
\multicolumn{1}{c}{(8)}&
\multicolumn{1}{c}{(9)}&
\multicolumn{1}{c}{(10)}\\
\multicolumn{1}{c}{Source}&
\multicolumn{1}{c}{$|$V$|$ (m/ms)}&
\multicolumn{1}{c}{V/V$_{\rm{t}}$--1}&
\multicolumn{1}{c}{$\psi $($\degr$)}&
\multicolumn{1}{c}{PA($\degr$)}&
\multicolumn{1}{c}{CA($\degr$)}&
\multicolumn{1}{c}{SNR}&
\multicolumn{1}{c}{Sep. (mas)}&
\multicolumn{1}{c}{Br. Ratio}&
\multicolumn{1}{c}{$\phi_{\rm UD}$ (mas)}\\
\hline 
SAO 164567                 & 0.7325 & 3\% & 7 & 78 & 14 & 49.2 & $ 8.4 \pm0.2$ & $ 6.8 \pm0.2$ & $  $ \\
SAO 110325                 & 0.8571 & $-0$\% & $-1$ & 59 & $-7$ & 37.0 & $ 7.8 \pm0.8$ & $ 13.4 \pm1.1$ & $  $ \\
SAO 80764                  & 0.6568 & $-3$\% & $-2$ & 73 & $-45$ & 26.3 & $ 42.5 \pm0.3$ & $ 14.9 \pm0.3$ & $  $ \\
SAO 185661                 & 0.3287 & $-5$\% & $-2$ & 155 & 60 & 23.7 & $ 37.9 \pm1.1$ & $ 19.3 \pm0.7$ & $  $ \\
IRC -30319 A-B   & 0.5647 & 3\% & 2  & 136 & 44 & 52.6 & $ 15.0 \pm0.1$ & $ 8.74 \pm0.04$ & $  $ \\
IRC -30319 B-C   &  &  &  &     &  & 16.1 & $ 21.8 \pm0.1$ & $ 1.98 \pm0.01$ & $  $ \\
17454891-2809333  & $0.7720$ & 4\% & 3 & 98 & 6 & 25.0 & $ 39.3 \pm0.7$ & $ 17.3 \pm0.9$ & $  $ \\
SAO 165154                 & $0.5870$ & 24\% & 14 & 117 & 62 & 6.2 & $ 43.0 \pm1.9$ & $ 4.7 \pm0.4$ & $  $ \\
RZ Ari    & 0.6520 & $-$2\% & 10 & 73   & 11  & 41.3 &     &    & $ 10.6\pm0.2 $ \\
SAO 76214 A-C        & $0.3500$ & $-5$\% & $-2$ & 131 & 56 & 7.8 & $ 13.0 \pm0.7$ & $ 2.4 \pm0.1$ & $  $ \\
IRAS 04395+2521            & 0.6301 & 11\% & 8 & 135 & 49 & 21.4 & $ 6.5 \pm0.2$ & $ 2.9 \pm0.1$ & $  $ \\
04440885+2540333  & 0.8013 & $-0$\% & $-0$ & 77 & $-10$ & 3.9 & $ 15.6 \pm0.8$ & $ 1.4 \pm0.1$ & $  $ \\
05415664+2707323           & 0.9208 & $-2$\% & $-3$ & 108 & 12 & 17.4 & $ 24.8 \pm0.3$ & $ 7.8 \pm0.3$ & $  $ \\
HD 283610                  & 0.5244 & $-5$\% & $-3$ & 121 & 38 & 9.1 & $ 19.4 \pm0.7$ & $ 6.1 \pm0.3$ & $  $ \\
04264187+2500314  & (0.8900) & $-$ & $-$ & (86) & (0) & 3.8 & $ 89.5 \pm1.0$ & $ 2.5 \pm0.1$ & $  $ \\
SAO 77000                  & 0.4995 & 2\% & $-2$ & 109 & 37 & 16.0 & $ 12.6 \pm0.3$ & $ 1.49\pm0.03$ & $  $ \\
\hline 
\hline 
\end{tabular}
\end{table*}

\subsection{\object{SAO 164567}}
This star was revealed as a binary in a Calar Alto observation
reported in F04. On the same night, the
star was observed also from TIRGO.
The binarity is clearly confirmed, but it is difficult to
extract a true position angle from the combination of the value
reported in F04 with that presented in
Table~\ref{table_results}. This is due partly to the relatively
small difference in position angle predicted for the two sites
(only $3\degr$), and partly to the fact that the Calar Alto event
was fitted with a speed that, in spite of just 2.9\% excess over
the predicted value, does not allow to compute unambiguously
the exact position angle. This happens occasionally when a LO event
has a very small contact angle. 

We can only conclude that the companion is generally oriented towards
the North, at a separation that could be significantly larger
than the projected value of Table~\ref{table_results}, up to
$\approx$50\,mas. Attempts to confirm the true position angle
by techniques such as speckle interferometry are possible.
From the two events we have reliable magnitude
differences both in the $R$ and the $K$ bands. This permits us
to infer that the secondary is bluer, by $R-K\approx1.5$\,mag,
than the primary. The primary is classified as a K5 giant
(Houk \& Smith-Moore \cite{houk88}), therefore we estimate that
the secondary should have $R-K\approx0.9$, which would
be consistent with a late A or early F star.

\subsection{\object{SAO 110325}}
This newly detected binary was the subject of several
previous observations by speckle interferometry
(McAlister \cite{mcalister78}, Hartkopf \& McAlister \cite{hartkopf84})
as well as LO (Evans \& Edwards \cite{evans81}). 
The star was reported as unresolved also by Hipparcos.
The fact that
none of these previous records
revealed the companion can be explained by the small
projected separation that we list in Table~\ref{table_results},
and possibly by the brightness ratio equivalent to $\Delta K=2.8$\,mag,
which might be even larger at shorter wavelenghts.

\subsection{\object{SAO 165154}}
A LO event for this star was reported by
Evans et al. (\cite{evans85}), who did not find evidence
of binarity. We note that the star is relatively faint in
the visual and the secondary might not have been detected
previously for reasons of dynamic range.

\subsection{\object{RZ Ari}}
The bright, O-rich M6 star 
\object{RZ Ari} (45~Ari, $\rho_2$~Ari, HR~867)
has been the subject of several investigations
by high angular resolution methods. 
Five previously available angular
diameter determinations are listed in the
CHARM2 catalogue (Richichi et al. \cite{CHARM2}).
The results are somehow heterogeneous,
including observations at various wavelengths in the
optical and near-IR by LO and LBI, and referring
to either uniform, partially or fully limb-darkened
disk diameters (UD, LD, FD respectively). 

The star is an irregular long-period variable, although 
the amplitude is relatively small (0.6\,mag,
Kukarkin et al. \cite{kukarkin71}). 
In the near-IR the amplitude of variability is
not well documented, and it can be assumed to be
even smaller. An examination of the data available
from the AAVSO shows a slight trend of increasing
luminosity by about 0.5\,mag over the past 30 years 
in which diameter measurements are available. 
Neglecting in a first approximation
significant changes of angular diameter due to
variability, we plot all available determinations
in Fig.\ref{figure_rzari}, using UD values.
The conversion from LD and FD to UD has been done
by using guidelines and conversion factors 
provided in the original
references. The uncertainties
in this conversion can be considered smaller than the
error bars on the diameter determinations.
It can be noted that there is a general agreement
among the various determinations. 
A weighted mean yields the UD value
$10.22\pm 0.12$\,mas.

\begin{figure}
\resizebox{\hsize}{!}{\includegraphics{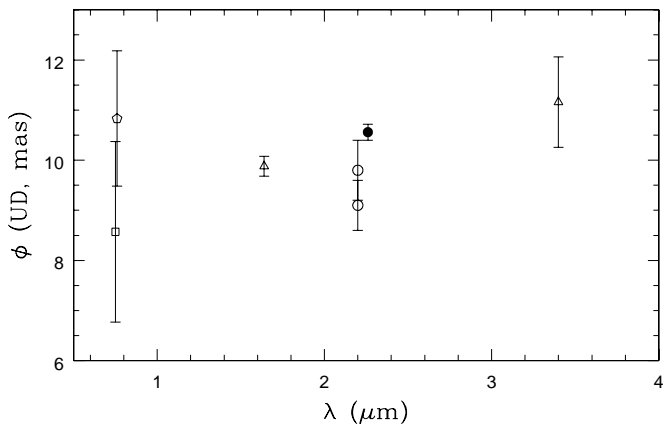}}
\caption{Angular diameter determinations for 
\object{RZ Ari}. The filled circle is our result, while
the open symbols are:
square Africano et al. (\cite{africano75}),
pentagon Beavers et al. (\cite{beavers81}),
triangles Ridgway et al. (\cite{ridgway80}),
circles Dyck et al. (\cite{dyck98}).
}
\label{figure_rzari}
\end{figure}

No definite trend of the characteristic size
with wavelength seems to be present, as 
would have been expected in the presence
of circumstellar matter, due to scattering at
shorter wavelengths and thermal emission at longer ones.
Therefore we can conclude that circumstellar matter
is not dominant. This is independently confirmed by
mid-infrared spectra, that show a featurless continuum
around 10$\mu$m (Speck et al. \cite{speck00}).
Also, there seems to be
no evidence of binarity, a possibility which had
initially been postulated on the basis of Hipparcos
results. Percy et al. (\cite{percy02}) have discussed
the origin of the problem with the Hipparcos data.
Also speckle interferometry investigations
by Mason et al. (\cite{mason99}) did not find
companions.
From our LO result, we can put an upper limit of
$\approx$1:40 on the brightness ratio of a hypothetical
companion with a projected separation in the range
$\pm 70$\,mas.

\object{RZ Ari} has been used as a building block
in several empirical T$_{\rm eff}$ calibrations, such
as those by 
Barnes (\cite{barnes76}, \cite{barnes78}),
Ridgway et al. (\cite{ridgway80}),
Di~Benedetto (\cite{dibene93}).
Dyck et al. (\cite{dyck98}) provided a revised value
of the bolometric flux, and using their own LBI diameter
derived 
T$_{\rm eff}=3442\pm148$\,K.
Of course, diameter variations must exist in this star,
and therefore it
seems of secondary importance at this point
to discuss the accuracy of the various determinations and
to refine the
T$_{\rm eff}$ value.
It would be more important to follow
diameter and temperature variations with a
dedicate monitoring, a possibility which is made
available by several of the current interferometers.

\subsection{\object{SAO 76214}}
Although this star is a known binary (Mason et al. 
\cite{mason01b}), 
our detection corresponds to a new component, with the
characteristics listed in Table~\ref{table_results}.
We detect also the
previously known component in our LO event record,
with a separation consistent with PA=$270\degr$ and
separation $0\farcs5$ listed by
Mason et al. 
(\cite{mason01b}), 
but it is outside the scope of our
observations to deal with such wide components.
Moreover, the quantitative evaluation of the trace
for \object{SAO 76214}
is hampered, especially on long time scales,
by significant scintillation.
We estimate the brightness ratio between the B
and the A-C components to be $0.56\pm0.10$
in the K band.
It is interesting to note that 
the Tycho Double Star 
Catalogue (Fabricius et al. \cite{tycho})
 examined this pair and found a $\Delta V$=2.59, however, LO
(Africano et al. (\cite{africano75})
and visual estimates (most recently, Worley \cite{worley89})
find a much smaller value, $\Delta V\approx 0.3$.

\subsection{\object{SAO 77000}}
This star has been repeatedly observed by filar micrometry
(Couteau \cite{couteau72, couteau75, couteau79,
couteau87, couteau89}, Heintz \cite{heintz80})
as well as by Hipparcos. Orbital motion is apparent over
the period of 20 years spanned by the observations, however
no clear orbital trend can be deduced yet. 
Due also to the intrinsically larger errors associated
with visual observations,
it is hard to extrapolate a possible position of the
component for the epoch of our LO event (2005.13). Nevertheless
we note a general consistency of quadrant and magnitude
of the separation.
Our measurement provides a significant constraint, since
it follows about 14 years after the most recent available
measurement.
Assuming to a first approximation that the
magnitude difference observed by Hipparcos
($\Delta$Hp=0.58\,mag) is similar to that
in the $V$ band, the comparison with the $K$ band
brightness ratio provided in 
Table~\ref{table_results}
indicates that the two components have almost the
same color, i.e. similar spectral types.

\subsection{Other binaries}
The remaining
stars listed in Table~\ref{table_results}
have no previous report of binary detection. Among these,
the following objects have at least
one bibliographical entry present in the {\it Simbad} database:
\object{SAO 80764},
\object{SAO 185661},
the triple star
\object{IRC -30319},
\object{IRAS 04395+2521}, and
\object{HD 283610}.
However these publications are on subjects not related
to high angular resolution observations.
There are no known previous publications associated 
with the four 2MASS objects present in 
Table~\ref{table_results}.

We also mention that we have detected binarity in three
further
stars from Table~\ref{table_list}, namely 
\object{SAO 109617},
\object{SAO 110089} and
\object{SAO 78540}. These are relatively wide systems,
with separations of order $0\farcs5$, and therefore easily
accessible to standard observations. For this reason, and
also because LO are not very accurate for such large
separations due to possible differences in local limb slope
for the two components, we have not included these results
in Table~\ref{table_results}. However, we consider it 
possibly useful to report the brightness ratios in the
$K$~band. The values are
$1.26\pm0.02$,
$1.70\pm0.03$ and
$0.5\pm0.1$,
in the above order.
It is noteworthy that all three stars have been measured
at visual wavelengths by speckle interferometry
and/or by Hipparcos.
We quote, among others, $\Delta$m values of
1.66\,mag ($G$ band, 
Balega et al. \cite{balega2004}) and
1.87\,mag (Hp band,
Fabricius \& Makarov \cite{fabr2000}) for
\object{SAO 109617},
and $\Delta$Hp values of
0.49\,mag and 1.73\,mag for
\object{SAO 110089} and
\object{SAO 78540}, respectively
(Fabricius \& Makarov \cite{fabr2000}).
We note that these latter authors provide also Tycho $B$
and $V$ magnitude differences.
We do not speculate at this point on the combination
of all these values with our $K$-band determination, in view
of the diversity of spectral bandpasses used in the visual.

A number of stars from Table~5 are additionally
wide binaries with separations of several arcseconds, and
we do not concern ourselves with them here.

\subsection{Other stars of interest}\label{other_interest}
Among the stars for which we did not detect binarity, a few are worthy
some comments either because of their nature or because of previous
attempts by high angular resolution techniques.
\object{SAO 80310} was investigated by Mason et al. 
(\cite{mason01a})
by speckle interferometry, with negative conclusions. 
The same result with the same technique was reported 
by Hartkopf \& McAlister (\cite{hartkopf84}) for \object{SAO 92659}. Both these stars were
also found
unresolved by Hipparcos.

\object{SAO 164601} is a spectroscopic binary, which was previously
observed as double by Evans et al. (\cite{evans86}). These authors
reported a separation close to 1\,mas, although without information
on the brightness ratio. We have analyzed our trace (SNR=18.7) with
both the LSM and CAL methods, without finding evidence of binarity.
In any case, due to the near-IR wavelength and the relatively
slow sampling, we are insensitive to separations of less than
about 3.5\,mas on this trace.
We notice that the position angle of our event ($110\degr$) was
almost orthogonal with that of the event observed
by Evans and collaborators.

We also recorded occultations during the passage of the
Moon over two regions of special interest.
On July 28th, 2004 the Moon reached a minimum distance
of $0\fdg59$ from the Galactic Center. In this crowded,
heavily obscured region
we could record 54 events at the 2.2\,m telescope in 3.4 hours,
being limited by overheads in telescope pointing and data storing.
The majority (50) of the objects has no counterpart in
optical catalogues. Spectral types on the other hand are known
for about half the sample, thanks mostly to the work of
Raharto et al. (\cite{rhi84}). With very few exceptions,
the stars are all of M spectral type.
From the photometry available in the 2MASS catalogue
(Cutri et al.~\cite{cutri03}), it can  be observed
that about half of the stars have a color $J-K>1$, indicating
significant reddening. This is presumably due to interstellar
dust in the direction of the Galactic Center, however in
some cases colors as red as 
$J-K=$ 3.5-5.0 are present, possibly pointing to
additional circumstellar extinction.

In January 2005, we were able to record
a passage of the Moon over the Taurus
star-forming region. These passages are
relatively frequent, and have been used in the past 
especially to derive important insights on the
frequency of binaries in the early stages of
stellar evolution (see Simon et al. \cite{simon95}, and
references therein). In our case, included were the
following known young stellar objects:
\object{LH 98-106},
\object{DL Tau},
\object{GN Tau},
\object{Elias 3-18},
\object{ITG 31},
\object{LkHA 332}. A few IR sources without
optical counterpart were also recorded.
Unfortunately, the sensitivity offered by the 1.5\,m telescope
was not sufficient to obtain quantitative results.
Details  on the full sample of occulted objects
can be found
in Table~5, available only on-line.

\section{Considerations on performance and statistics}\label{performance}
In F04 we reported
the limiting sensitivity computed
for LO observations with the MAGIC instrument. 
It was shown that the logarithm of the SNR of a LO light curve
is approximately in inverse linear relation to the K magnitude.
At the 1.5\,m telescope, with the typical integration and
sampling times of 3 and $\approx$8\,ms respectively,
it can be expected to detect objects having $K\approx8$\,mag
with SNR=3. 
The present sample is much larger than that of
F04.
Excluding
the sources observed from
TIRGO (since those observations employed an entirely
different detector and a comprehensive statistics for that
configuration was already provided by 
Richichi et al. \cite{richichi96}), 
\object{RZ Ari} which was observed with a narrow-band
filter, and a number of sources which were deemed too faint
and plainly not binary and consequently without a detailed
analysis,
we are left with 285 events.
However, the SNR-$K$ relationship of
the present sample  is not straighforward to interpret. 
Firstly,
about 20\% of the stars were observed with the 2.2\,m
telescope and show a trend which is offset from the
main relationship by the expected factor of mirror area.
Secondly,  about 2/3 of the runs at the 1.5\,m telescope
were carried out with a wrong position of the pupil wheel
which holds the cold stop. This had no effect on the
stellar signal, but has produced a large increase
in thermal background, resulting in higher noise, 
resulting in lower SNR than expected for a given
stellar magnitude. 
As a consequence our sample is more 
inhomogeneous than that of 
F04, although the general
characteristics of the relationship are confirmed.
At the 1.5\,m telescope we recorded about 20 events
for stars with $K$ between 8 and 8.5\,mag.
At the 2.2\,m telescope, used
only for the very crowded passage near the Galactic
Center, we had a sufficient number of bright sources and the
real limiting magnitude was not reached.

Similarly to what was done in F04,
we have computed also
the limiting angular resolution associated with the
unresolved sources,
following the same approach of Richichi et al.
(\cite{richichi96}). 
This has been done for 103 stars in our sample having
SNR$>$10, or
about 4 times more numerous than in the sample of
F04. The result, including a comparison
with this latter work, is shown
in Fig.~\ref{limres}.

\begin{figure}
\resizebox{\hsize}{!}{\includegraphics{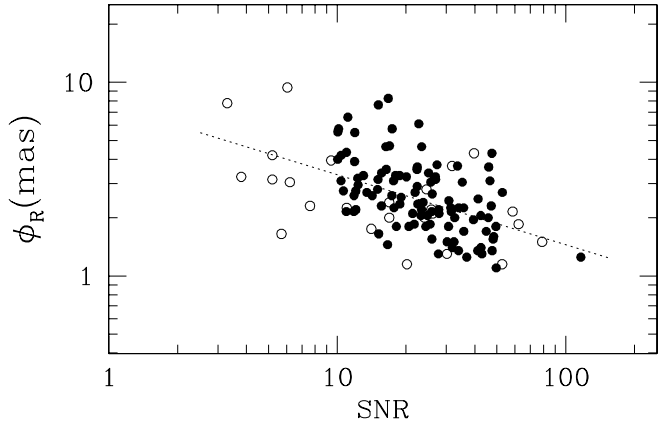}}
\caption{Limiting resolution 
for the sources in our sample, as a function
of SNR (solid dots).  Only points with SNR$>$10
are included. Also shown as open circles
are the determinations from Fors et al. (\cite{fors04}).
The solid line is a log-log fit through all points.
}
\label{limres}
\end{figure}

It can be noted that the current sample and the previous
from F04
have an almost identical
distribution of limiting resolution against SNR, 
and can be fitted by the same log-log relationship.
This is reassuring, since the behaviour
be independent
of the source, and be determined by the instrumental
characteristics and in particular by the integration time.
The large spread in the relationship can be understood
in terms of large variations of SNR from one LO light curve
to another due to different situations of background
and also to the specific conditions
 of signal extraction from the
discrete pixels of the detector.
Broadly speaking, the average relationship is such
that SNR=10 ensures a limiting resolution of about
3\,mas. In the few cases in which SNR 
close to 100
could be recorded, the limiting resolution improves but
remains above, as already noted 
in F04, the performance of
fast InSb photometers which can operate with faster
sampling.

A final consideration can be made about the statistics
of binary detections in our sample. We have observed
a total of 14 binaries (counting as such also the triple
star \object{IRC -30319}), out of a total sample
size of 350 stars. This points to a fraction of
4.0\%, or more than two times smaller than what
observed by
Richichi et al. (\cite{richichi96}) and in F04.
This result seemed puzzling at first, since all the
samples considered have a broad sky distribution
and should have similar characteristics. 
It is not excluded that the targets that we observed
in the direction of the Galactic Center have
an actual deficit of binaries, due to the fact
extinction introduced a bias
towards stars
that for a given apparent magnitude are more distant
than in the previous samples.
Therefore, hypothetical companions would have smaller 
angular separations for the same statistics of semi-major
axis. However, only 20\% of the stars in our sample
were observed in the direction of the Galactic Center,
and another explanation must exist for the lower
binary fraction that we observe in the present work.

In fact we note that with the introduction of large, deep
IR catalogues such as 2MASS in our predictions, we have
effectively shifted the distribution of $K$ magnitudes
in our sample much closer to the limiting sensitivity
of the technique. Therefore, we can expect that most
of the LO light curves will have on average lower
SNR than in the previous samples.
As a result, it will become effectively more difficult
to detect companions, especially those with brightness
ratios larger than unity. Although we have not performed
a detailed computation of this effect, its magnitude
could easily explain the observed apparent deficit 
of binary detections. We conclude that the introduction
of large catalogues, while increasing the number
of predictions and correspondingly of observed LO,
does not automatically produce a higher rate of results.

\section{Conclusions}
We have provided an update on the 
program of lunar occultation observations which is
operational at 
Calar Alto Observatory, previously described by
Fors et  al. (\cite{fors04}), including
350 lunar occultation events. Although no major
changes have occurred with respect to instrumentation,
the program has been expanded to include, in addition to 
the Spanish 1.5\,m telescope, also the 2.2\,m telescope.
Additionally, we have developed and made use of
new methods of 
light curve
extraction and characterization, suitable
to perform in an automated fashion the
preliminary analysis of large volumes of lunar occultation data.
This has been made necessary by the availability of large, deep
near-IR catalogues such as the 
2MASS (Cutri et al.~\cite{cutri03}) 
and DENIS (Paturel et al.~\cite{paturel03}), which permit
the prediction and observation 
of a much increased number of occultation events.

The results include the detection in the near-IR of
one triple and
13 binary systems.
For all but two stars, these represent first
time detections.  Projected separations range
from $0\farcs09$ to $0\farcs007$, and brightess
ratios reach up to 1:20 in the $K$ band.
We have also determined the angular diameter of
the M6 star 
\object{RZ Ari}, which we have discussed in
comparison with previous determinations.
Our observations have included a passage of the
Moon over a crowded region in the vicinity of
the Galactic Center (resulting in 54 events observed
in about 3 hours), and a passage in the Taurus
star-forming region.
Passages of the Moon close to the Galactic center
are taking place in these years, and we have provided
some examples. These events provide a unique
opportunity to extract milliarcsecond resolution
information on a large number of objects in obscured,
crowded and relatively unstudied regions, and can
be adequately observed with 2-4\,m -class telescopes.

We have discussed the performance achieved in our
observations in terms of limiting magnitude and
angular resolution. We have shown that at 1-2\,m class
telescopes equipped with a rather traditional
array detector it is possible
to achieve $\approx 0\farcs003$ on sources as faint
as $K\approx 8$\,mag. The rate of binary detection
in random observations of field stars that emerges
from the present work is $\approx 4$\%, considerably
lower than established earlier by similar 
studies
(Richichi et al. \cite{richichi96},
Fors et al. \cite{fors04}).
We attribute this effect largely to the fact that the use
of catalogues such as 2MASS has increased dramatically
the number of occultation observable per night, but this
increase is realized mostly at the faint magnitude
end, where the dynamic range available is much
smaller than for brighter stars.

\begin{acknowledgements}
We thank the Observatorio Astron\'omico Nacional (OAN) and  Centro Astron\'omico
Hispano-Aleman (CAHA) for the facilities and  support made available at Calar
Alto, and in particular Santos Pedraz (CAHA) for his
invaluable help on many occasions. This publication makes use of 
data products from the \textit{Two Micron All Sky Survey}, which is a  joint
project of the University of Massachusetts and the Infrared  Processing and
Analysis Center/California Institute of Technology,  funded by the National
Aeronautics and Space Administration and the  National Science Foundation.  We
acknowledge with thanks the variable star observations from the  AAVSO
International Database contributed by observers worldwide and  used in this
research. This research has made use of  the \textit{Simbad} database, operated
at CDS, Strasbourg (France), and of the Washington
Double Star Catalog maintained at the U.S. Naval Observatory.
The paper has been improved by
several detailed references provided by the referee, Dr. B. Mason.
\end{acknowledgements}

\end{document}